# Dual-Mode Exciton Coupling in Epitaxially Registered Organic-Inorganic 2D Heterocrystals

Eunbeen Jeon[1], Kihyun Lee[2], Jieun Yeon[2], Juseung Oh[1], Kenji Watanabe[3], Takashi Taniguchi[4], Kwanpyo Kim[2] and Sunmin Ryu[1]*

[1]Department of Chemistry, Pohang University of Science and Technology (POSTECH), Pohang, Gyeongbuk 37673, Korea

[2]Department of Physics, Yonsei University, Seoul 03722, Korea

[3]Research Center for Electronic and Optical Materials, National Institute for Materials Science, 1-1 Namiki, Tsukuba 305-0044, Japan.

[4]Research Center for Materials Nanoarchitectonics, National Institute for Materials Science, 1-1 Namiki, Tsukuba 305-0044, Japan.

*E-mail: sunryu@postech.ac.kr

**Abstract**

Two-dimensional (2D) heterocrystals comprising molecules and semiconductors can serve as an ideal platform for studying interfacial excitons and for future optoelectronic applications, yet the energy and charge flow across these atomically sharp interfaces remain unclear. In this work, we investigated PTCDA/$MoS_2$ as a prototypical 2D organic-inorganic heterostructure and revealed dual-mode exciton coupling between the constituent crystals. Monolayer-resolved PTCDA molecular crystals were grown on monolayer $MoS_2$ via physical vapor assembly, and their crystallographic details, including the stacking angle, were determined by electron diffraction. Upon the formation of the heterostructures, PTCDA's photoluminescence was nearly completely quenched by efficient interfacial excited-state decay, whereas that of $MoS_2$ increased markedly with PTCDA thickness. Using differential reflectance and photoluminescence excitation (PLE) spectroscopy, we found that the enhancement arises from two distinct contributions. One contribution, observed without direct PTCDA excitation, is consistent with ground-state charge transfer injecting holes into $MoS_2$, which suppresses negative trion formation and enhances the radiative recombination of neutral excitons. The other, correlated with PTCDA absorption in the PLE response, supports resonant energy transfer, which diverts excitation energy from PTCDA to $MoS_2$. Our findings help reconcile previously proposed mechanisms and show how molecular and $MoS_2$ excitonic responses evolve with molecular thickness in a structurally defined heterocrystal.

## Introduction

Two-dimensional molecular crystals (2DMCs) provide an ideal platform for investigating excitonic phenomena governed by intermolecular interactions[1] under reduced dimensionality. Their atomically thin geometry enables control over thickness[2-4], dielectric screening[5], and epitaxial registry[6, 7], allowing systematic tuning of electronic structure and excitonic coupling. Previous studies have shown that high-quality 2DMCs exhibit long-range order[2, 7], optical anisotropy[4], and Davydov splitting[4, 8], establishing polarization-resolved spectroscopy as a direct probe of molecular packing and transition dipoles[4, 8, 9]. Notably, molecular excitons in monolayer (1L) 2DMCs are highly delocalized, showing room-temperature superradiance[10, 11] and supertransport[3] behavior. The excitonic structure also evolves with thickness: excitations in perylene systems, for example, are predominantly Frenkel-like in the monolayer limit, whereas increasing thickness introduces charge-transfer (CT) character through intermolecular π-stacking, leading to Frenkel-CT mixing[8, 10] and reconfiguration of transition dipoles[8].

While these studies establish the intrinsic excitonic properties of 2DMCs, parallel efforts have extended their functionality by interfacing them with photoactive 2D semiconductors. Hybrid organic-inorganic heterostructures[7, 9, 12] combine strong molecular absorption with efficient excitonic emission in transition metal dichalcogenides (TMDs), providing a platform for engineering interfacial exciton dynamics[13-19]. Molecular orientation and packing depend sensitively on substrate interactions, leading to face-on or edge-on configurations and varying degrees of crystallinity[20]. Mechanistic studies have demonstrated charge transfer[16], energy transfer[14, 17], and band alignment effects[17] in organic/TMD heterostructures, enabling functionalities such as vertical transport[21], hybrid p-n junctions[22], and phototransistors[23]. Despite these advances, existing studies have largely focused on several-nanometer-thick or polycrystalline molecular films deposited on inorganic 2D semiconductors[16, 17, 21, 22], where variations in molecular morphology and grain structure can complicate the interfacial response and preclude monolayer-resolved thickness control. In addition, thicker molecular films can support excimer formation, introducing additional excited-state pathways and further modifying the interfacial response. Whereas recent work has demonstrated control of optical anisotropy and charge localization in monolayer molecular crystals on TMDs[9, 12], the evolution of intermolecular coupling along the out-of-plane direction and its interplay with interfacial processes remain unclear.

In this work, we grow thickness-controlled 2D PTCDA crystals on 1L $MoS_2$ with epitaxial registry and reveal dual-mode excitonic coupling in the prototypical 2D organic-inorganic heterostructure. Our monolayer-resolved heterocrystals provide a structurally well-defined interface with controlled molecular thickness, order and registry. Electron diffraction confirms orientational registry, and polarization-resolved spectroscopy reveals long-range molecular orientational order at the molecule-TMD interface. Davydov-split molecular excitons identified by polarization-resolved absorption spectroscopy are strongly coupled to $MoS_2$, as evidenced by near-complete quenching of photoluminescence (PL). In contrast, excitonic emission from $MoS_2$ exhibits enhancement consistent with both ground-state charge-transfer doping and resonant energy transfer. We also reveal that the charge-state-sensitive response of $MoS_2$ continues to evolve beyond the first PTCDA layer, suggesting electronic communication through the out-of-plane π-stacked molecular layers. Layer-resolved measurements further show that PTCDA excitonic renormalization is strongest at the immediate interface, whereas $MoS_2$ exciton-trion balance and PL response continue to evolve with increasing PTCDA thickness. PL and PL excitation measurements at different excitation energies distinguish a response independent of molecular excitation from an additional contribution upon PTCDA excitation.

## Results and Discussion

***Formation and excitonic basis of PTCDA/$MoS_2$ heterocrystals.*** The transient flow of photoexcitation across atomically sharp interfaces may proceed through resonance energy transfer (RET) or charge transfer (CT), as illustrated in Fig. 1a. To establish a well-defined platform for interfacial exciton coupling, we first characterize the structural and excitonic properties of thickness-controlled PTCDA/$MoS_2$ heterocrystals. Using physical vapor assembly, uniform single- and few-layer PTCDA crystals (denoted as $nL_P$; n = 1–5) were grown on monolayer $MoS_2$ ($1L_M$). Based on the systematic characterization that follows, we found that $nL_P$ is essentially a herringbone-patterned array of 1D π-stacked PTCDA columns (Fig. 1a). Figure 1b (middle) shows an AFM image of an edge region of a partially grown $2L_P$ sample, which was selected to show

PTCDA-$MoS_2$ steps. The majority of the grown samples were completely and uniformly covered with molecular layers. The step height of 0.62 ± 0.12 nm (Fig. 1b, top) corresponds to two-layer thickness based on the interplanar spacing of the (102) plane in bulk PTCDA[24]. These results suggest layer-by-layer growth with crystallographic fidelity. The narrow height distribution (Fig. 1b, bottom) further indicates high surface quality.

Optical transitions in the organic-inorganic heterocrystal are fundamentally different in the two constituents. Whereas optical absorption in 1L $MoS_2$ originates from direct interband transitions between delocalized Bloch states near the K valleys[25, 26], photoexcitation in PTCDA involves transitions between localized molecular orbitals accompanied by vibronic progressions (Fig. 1c). This fundamental contrast provides an ideal platform for exploring interfacial energy and charge flow between molecular excitons and delocalized band excitons. To characterize the electronic structure of the heterocrystals, we employed differential reflectance (DR) spectroscopy: the absorptance of a thin sample supported on a dielectric substrate is proportional to DR, defined as $(R_{sam} - R_{sub})/R_{sub}$, where $R_{sam}$ and $R_{sub}$ are the reflectances of the sample and substrate areas, respectively[27]. As shown in Fig. 1d, the unpolarized DR spectrum of $3L_P/1L_M$ exhibits both the A, B and C excitons of an isolated $1L_M$ (denoted as $X_A$, $X_B$ and $X_C$)[25, 26] and two vibronic peaks of an isolated $3L_P$ grown on hexagonal BN (hBN)[8], assigned to the $S_1 \leftarrow S_0$ transitions (denoted as 0-0 and 1-0)[28]. The characteristic excitonic resonances of both constituents remain clearly identifiable in the heterocrystals, indicating that their fundamental excitonic structures are largely retained upon heterocrystal formation. However, the $MoS_2$ A exciton blueshifts whereas the molecular resonances redshift, indicating non-negligible interlayer coupling. The thickness dependence of the spectral changes is also different for the two constituents: with increasing thickness, $X_A$ and $X_B$ blueshift, whereas the molecular peaks remain largely constant (Fig. 1e). Note that the spectra in Fig. 1e are shown with baseline subtracted (see Fig. S1 for detailed analysis). The origin of these spectral changes will be discussed below using polarization-resolved measurements. Quantitative analysis in Fig. 1f shows that molecular absorption scales linearly with thickness, enabling optical determination of the number of layers. It should be noted that the optical contrast[8] obtained from optical micrographs misidentifies thickness due to annealing effects on $MoS_2$'s optical response (Fig. S2).

***Lattice structure and epitaxial registry of heterocrystals.*** The layer-by-layer growth suggests long-range order and epitaxial registry. Electron diffraction of $2L_P/1L_M$ (Fig. 2a) reveals well-defined rectangular molecular lattices, which were verified across multiple spots (5 µm apart) in multiple samples (Table S1). As found for PTCDA grown on hBN and graphene[8], the molecular lattices correspond to the (102) plane of bulk alpha-type crystals[24]. However, quantitative analysis revealed that the 2D unit cell spanned by $\boldsymbol{m}$ and $\boldsymbol{n}$ unit vectors is significantly dependent on the assembly templates: Whereas $\boldsymbol{m}$ of $1L_P$ is 4.2% lengthened on hBN[8], $\boldsymbol{m}$ ($\boldsymbol{n}$) of $2L_P$ is 6.7% (1.0%) lengthened (shortened) on $MoS_2$. Low-magnification diffraction patterns further reveal a reproducible relative orientation between the PTCDA and $MoS_2$ lattices (data are not shown): the reciprocal primitive vectors ($\boldsymbol{a}^*$ and $\boldsymbol{b}^*$) of the hexagonal lattice of $1L_M$ are given with $\boldsymbol{m}^*$ and $\boldsymbol{n}^*$ of $2L_P$ (Fig. 2a). The stacking angle, defined by the angle between $\boldsymbol{n}^*$ and $\boldsymbol{b}^*$, was found to be constant (~13°) for multiple samples of $2L_P$ and $4L_P$ (Table S1), which indicates preferential orientation of the molecular lattice on $MoS_2$. Noting that the hexagonal lattice of $MoS_2$ is 26% larger than that of hBN or graphene, these findings indicate that the molecule-template epitaxial registry is an important factor governing molecular packing structure. With increasing PTCDA thickness to 4L, $\boldsymbol{m}$ decreased to within ~2% of the bulk value, further suggesting 2D-to-3D convergence[8].

Using polarization-resolved DR measurements, we confirm the presence of two basis molecules in the unit cell, which is consistent with the (102) plane of bulk crystals. Coulomb interaction between two transition dipoles localized at two basis molecules leads to two superposition states, lower (symmetric) and upper (antisymmetric) Davydov states (LDS and UDS), as depicted in Fig. 2b. Unlike the unpolarized DR spectra in Fig. 1, the parallel-configuration spectra ($DR^{\parallel}$ defined in Methods) of $2L_P$ for selected orientation angles (Fig. 2c) show that the 0-0 absorption consists of two subcomponents due to the Davydov splitting, a hallmark of multiple bases[29]. As shown in the full-set spectra (Fig. S3), their contributions vary systematically with sample orientation, arising because the two Davydov states are orthogonally polarized (Fig. 2d). As shown in Fig. S4, the same two components with nearly orthogonal transition dipoles are resolved in $1L_P$, excluding an origin from two mutually rotated molecular layers. This near-orthogonal relationship is consistently observed in 29 heterocrystals spanning 1–$5L_P$ (Fig. S4), consistent with Kasha's dimeric exciton model[4, 30], which we use here as a phenomenological description of the Davydov pair. In extended molecular crystals and organic/TMD interfaces, however, exciton delocalization, long-range Coulomb coupling, dielectric screening and CT or hybrid excitations can modify the excited-state character beyond this simple picture[31-33]. The dipolar character of the molecular transitions contrasts with the isotropic responses of $MoS_2$, which originate from its threefold rotational symmetry (Fig. 2d). We also note that the 1-0 absorption reveals a slight but similar energy splitting (Fig. S3).

***Static interlayer coupling: dielectric screening and charge-transfer doping.*** We next examine static interlayer coupling manifested in absorption spectra and exciton energies. Evidence of interfacial electronic coupling is provided by the ~100 meV stabilization and broadening of the 0-0 LDS of $1L_P/1L_M$ relative to the $1L_P$/hBN reference (Fig. 3a). Assuming negligible electronic coupling of PTCDA with dielectric hBN[8, 34], the observed redshift and broadening on $MoS_2$ may result from dielectric screening[35] or wavefunction mixing[16]. First of all, the redshift of the 0-0 transition was observed for LDS and UDS of all thicknesses, as shown in Fig. 3b. With decreasing thickness from $5L_P$ to $2L_P$, both 0-0 transitions blueshift because of reduced polarization stabilization in thinner layers[36]. Then, a drastic redshift for $1L_P$, ~30 meV (LDS) and 37 meV (UDS), indicates a significant interaction between $MoS_2$ and the neighboring molecular layer. We attribute the coupling to interfacial dielectric screening by $MoS_2$[35] relative to hBN, reflecting its larger dielectric response and electronic polarizability. This interpretation is also consistent with an electrostatic screening model that predicts renormalization of molecular electronic levels at an organic/TMD interface[37]. Previous studies of PTCDA/$MoS_2$ have reached different conclusions regarding interfacial electronic hybridization: DFT calculations suggested orbital mixing at the interface[16], whereas photoelectron spectroscopy found no clear evidence for substantial hybridization[17]. Thus, we regard dielectric screening as a well-supported contribution to the observed spectral renormalization, while an additional contribution from interfacial wavefunction mixing cannot be excluded or quantified from the present optical measurements. Notably, the near-orthogonality of the LDS and UDS transition dipoles remains preserved despite these interfacial interactions (Fig. 2d). The linewidths of LDS and UDS of $1L_P$ increased by ~90% and ~40%, respectively, compared to the reference (Fig. 3c). The additional broadening may contain both inhomogeneous and homogeneous contributions, including exciton-phonon scattering and fast interfacial decay. As will be shown by the drastic quenching of molecular PL below, the organic-inorganic interlayer coupling opens a highly efficient nonradiative decay channel[15] via hole transfer, which may contribute to the observed linewidth broadening. The noticeable reduction in DS energy relative to that of the reference is also due to interlayer interactions (Fig. 3d). Interestingly, the degree of the splitting is maintained for all thicknesses.

A complementary insight into the organic-inorganic coupling is provided by the DR response of $MoS_2$ excitons, which blueshift upon formation of heterocrystals. The gradual increase of ~15 meV in $X_A$ energy with increasing thickness (Fig. 3b) is due to the interplay between neutral and charged A excitons. PTCDA molecules with a large electron affinity (~4.12 eV) have a strong tendency to withdraw electrons from (or inject holes into) $MoS_2$[16, 18], which has significant native negative charges[38]. As shown by the electrical gating experiments[38], hole doping diminishes the absorption of the predominant negatively charged trions and increases that of the neutral excitons. Because the two peaks are barely resolved due to their small energy difference, the change appears as an apparent upshift, which is in fact a population shift towards high-energy neutral excitons. The PTCDA-induced narrowing of $X_A$ in Fig. 3c is also consistent with this interpretation. In other words, the large linewidth of bare $MoS_2$ is due to the contributions from neutral excitons and trions.

PL spectra of heterocrystals further reveal PTCDA-induced charge-transfer doping of $MoS_2$. Compared with bare $MoS_2$, the $X_A$ of heterocrystals blueshifts and increases in intensity (Fig. 4a). The exciton energy (Fig. 4b) and PL intensity (Fig. 4c) increased systematically with PTCDA thickness, a trend consistently observed for 514 and 633 nm excitation (Figs. 4b and 4c). These thickness-dependent trends were reproduced across multiple heterocrystals, with the number of samples examined for each optical measurement summarized in Table S2. Spatially resolved DR and PL line scans of a representative $1L_P/1L_M$ heterocrystal also show only small variations in the molecular and $MoS_2$ spectral parameters over the probed region (Fig. S5). Notably, the change in exciton energy was essentially equivalent for the two wavelengths. Because 633 nm excitation lies below the optical gap of PTCDA (i.e., sub-gap excitation), we conclude that the blueshift does not require molecular excitation but results from ground-state PTCDA via charge-transfer doping[18, 39]. To obtain a semi-quantitative estimate of the electron density reduced by PTCDA, we compared the trion fraction in $X_A$'s PL intensity with the gate-dependent calibration reported by Mak *et al.* for monolayer $MoS_2$[38]. The trion fraction is obtained directly from the PL spectral decomposition, whereas its conversion to electron density assumes approximate transferability of this literature calibration to the present heterocrystal systems. Thus, the resulting carrier densities should be regarded as approximate estimates. As shown for 1L in Fig. 4a, $X_A$ of all thicknesses could be decomposed into negative trion ($X_A^-$) and neutral exciton ($X_A^0$) subcomponents (Fig. 4d). With increasing thickness, the trion fraction decreased (Fig. 4e). According to the calibration data in Fig. S6, our bare $MoS_2$ ($0L_P$) corresponds to an electron density ($n_e$) of $8.2 \times 10^{12}$ $cm^{-2}$, whereas 1L PTCDA/$MoS_2$ corresponds to $n_e = 7.4 \times 10^{12}$ $cm^{-2}$. Thus, deposition of a single PTCDA layer is estimated to remove $8 \times 10^{11}$ electrons/$cm^2$. This corresponds to approximately 0.01 electrons per PTCDA molecule, given the molecular packing density of 1L PTCDA ($8.7 \times 10^{13}$ molecules/$cm^2$; see Table S1).

It is notable that the charge-state-sensitive optical response of $MoS_2$ continues to evolve beyond the PTCDA layer directly adjacent to $MoS_2$: A similar estimate in Fig. S6 reveals that 5L PTCDA corresponds to an approximately twofold larger reduction in electron density than 1L PTCDA. Because this estimate is derived from the optical trion fraction, its quantitative value may also be influenced by thickness-dependent dielectric screening and the local electrostatic environment, which can modify the exciton-trion oscillator strengths, linewidths and radiative efficiencies. Nevertheless, the systematic decrease in the measured trion fraction with increasing PTCDA thickness indicates that molecular layers beyond the immediate interface influence the charge state of $MoS_2$. Such a thickness-dependent influence may be facilitated by electronic coupling along the out-of-plane direction, consistent with the π-stacked 1D columnar structure of PTCDA crystals[40]. Such molecular stacking is expected to provide electronic communication between successive PTCDA layers, although the microscopic origin of the thickness-dependent charge redistribution is not directly resolved in this work. $X_A$'s PL enhancement for the 633 nm excitation in Fig. 4c also agrees with the static charge-transfer doping: with increasing thickness, the residual electron density of $MoS_2$ decreases due to the doping, and $X_A$ state is more populated by neutral excitons ($X_A^0$) with higher PL quantum yield than negative trions ($X_A^-$)[41, 42], as illustrated in Fig. 4f. To avoid complications arising from annealing-induced PL modulation[42], the PL enhancement was referenced to bare 1L $MoS_2$ annealed at the growth temperature.

***Dynamic interlayer coupling: photoluminescence modulation.*** The PL response of the heterocrystals is also substantially modified by excited-state interlayer coupling. The prominent molecular PL is reduced by a quenching factor of ~500 for $1L_P/1L_M$ (Fig. 4a). In contrast, the $X_A$ emission is enhanced with increasing thickness for the 514 nm excitation (Fig. 4c). Notably, the enhancement is twice as large as that for the 633 nm excitation: for the maximum thickness of $5L_P$, the steady-state coupling caused a fivefold enhancement for 633 nm, and the dynamic coupling induced an additional twofold increase for 514 nm. To shed light on the origin of the additional PL enhancement by the above-gap excitation, we obtained PL excitation (PLE) spectra for $1L_P$ and $4L_P$ in Fig. 5a. The PLE signals, given in the form of the PL enhancement, closely follow the DR spectra, suggesting that photoexcitation is transferred from PTCDA to $MoS_2$.

We now address the microscopic origin of dynamic interlayer coupling, which underlies the broadening of the 0-0 absorptions, the drastic quenching of molecular PL, and the enhancement of $X_A$ PL. The substantial spectral overlap between the PTCDA PL and the $MoS_2$ absorption (Fig. 5b), together with the in-plane alignment of their transition dipoles[8], provides favorable conditions for efficient RET. The close correspondence between the PLE and molecular absorption spectra (Fig. 5a) further supports the involvement of RET in the additional PL enhancement, as depicted in Fig. 5c. Park et al. also proposed efficient RET from PTCDA to $MoS_2$ to explain a sub-ps decay in the absorption of excited PTCDA[17]. In parallel with RET, interfacial resonance-controlled nonadiabatic CT[15] also contributes to the excited-state dynamics. The highest occupied molecular orbital (HOMO) of PTCDA is 0.5 eV lower than the valence band (VB) maximum of $MoS_2$, providing a favorable downhill driving force for hole transfer. Combined with the interfacial electronic coupling, this energetic alignment promotes efficient resonance-controlled nonadiabatic hole transfer from PTCDA to $MoS_2$ (Fig. 5d), which can account for the drastic quenching of the molecular PL. Such hole transfer may also contribute to the ~35 meV broadening of the LDS absorption in Fig. 3c. As a limiting estimate, if the entire additional linewidth broadening is treated as a lifetime-limited Lorentzian homogeneous contribution from interfacial hole transfer, the 35-meV increase corresponds to a timescale of ~20 fs. Because inhomogeneous broadening and exciton–phonon scattering may also contribute to the observed linewidth, this value should be viewed as a model-dependent estimate compatible with rapid interfacial hole transfer, as reported in other 2D heterostructures[43]. A minimal kinetic framework for these competing excited-state pathways is provided in the Supplementary Note, where RET and photoinduced charge transfer are treated as parallel decay channels of photoexcited PTCDA. The present steady-state measurements, however, do not quantify the relative contributions of RET and photoinduced CT.

Overall, our PTCDA-$MoS_2$ system exhibits type-I-like exciton dynamics, in which photoinjected energy is funneled into $MoS_2$, which has a smaller bandgap. We note, however, that nonadiabatic interfacial CT also competes with RET for the photoexcited PTCDA population. For the above-gap excitation, hole transfer can divert excitonic energy in PTCDA toward $MoS_2$ to form interlayer CT excitons (Fig. 5d). For the sub-gap excitation, the role of nonadiabatic electron transfer from $MoS_2$ to PTCDA is intriguing. Because the PTCDA LUMO and $MoS_2$ CBM are nearly resonant[17], forward electron transfer from photoexcited $MoS_2$ to PTCDA, followed by back electron transfer to $MoS_2$, may provide a reversible excited-state pathway (Figs. 5e and 5f). Such a reversible transfer pathway could limit persistent interfacial charge separation, although the individual transfer steps are not directly resolved in the present measurements. As the lifetime of $X_A$ is ultrashort[44], this pathway would not necessarily lead to substantial quenching of the $X_A$ emission. Nevertheless, the PL enhancement induced by the sub-gap excitation (Fig. 4c) is consistent with ground-state charge-transfer doping (Fig. 4f), for which RET is not operative.

Finally, we connect the current work with previous studies on several-nm-thick polycrystalline PTCDA films/$MoS_2$. Using two-photon photoemission, Rijal et al. observed ultrafast electron transfer from $MoS_2$ to PTCDA, followed by the formation of interlayer excitons[15]. Despite the technical difficulty posed by their low oscillator strength, PL detection of interlayer excitons warrants experimental effort, as it will complete the picture of exciton dynamics in this heterosystem. Park et al. characterized a type-I interfacial electronic structure using angle-resolved photoemission and observed 100% PL enhancement[17], which is consistent with the current work. Whereas Obaidulla et al. assumed a type-II structure[16], they observed enhanced PL, which was attributed to multiple origins including static charge transfer doping, orbital hybridization and morphology-dependent nonradiative decay. It should be noted that the thick polycrystalline films employed in those studies can host molecular excimers and provide dielectric and electrostatic environments that differ substantially from those experienced by the few-layer epitaxial crystals investigated here. In contrast, the present monolayer-resolved, epitaxially registered PTCDA crystals provide a more structurally defined interface and enable the molecular and $MoS_2$ responses to be followed systematically with PTCDA thickness.

## Conclusions

In this work, we investigated thickness-controlled PTCDA/$MoS_2$ heterocrystals and established a comprehensive picture of their structural and electronic properties. Electron diffraction combined with polarization-resolved spectroscopy revealed long-range molecular orientational order, epitaxial registry, and Davydov-split molecular excitons, demonstrating that the organic layer forms a structurally and excitonically well-defined interface with $MoS_2$. Systematic optical measurements showed that interlayer coupling manifests through both spectral shifts and linewidth broadening, reflecting dielectric screening and the emergence of efficient nonradiative decay channels. In parallel, $MoS_2$ excitons exhibit thickness-dependent energy shifts and linewidth narrowing, which we attribute to hole injection induced by ground-state charge transfer from PTCDA. The layer-resolved measurements further show that PTCDA excitonic renormalization is strongest at the immediate interface, whereas $MoS_2$ exciton-trion balance and PL response continue to evolve with increasing PTCDA thickness. Excitation-dependent photoluminescence distinguishes a response independent of molecular excitation from an additional contribution upon PTCDA excitation. The former is consistent with ground-state charge-transfer doping that modifies exciton populations in $MoS_2$, whereas the latter follows PTCDA absorption. Photoluminescence excitation measurements further support the involvement of resonance energy transfer in this additional enhancement. Overall, our results demonstrate that excitonic coupling in organic-inorganic heterostructures can be systematically tuned by molecular thickness. These findings highlight the value of structurally defined heterocrystals for resolving interfacial excitonic interactions in organic-inorganic 2D systems.

## Methods

***Preparation of 2D heterocrystals***. Single and few-layer PTCDA molecular crystals were formed by physical vapor assembly using 1L $MoS_2$ and few-layer hBN as assembly templates, as reported previously[8]. Briefly, the inorganic 2D crystals were prepared by mechanically exfoliating[45] their bulk counterparts (2D Semiconductors Inc., flux-grown $MoS_2$) onto amorphous quartz substrates (SPI, SuperSmooth), unless otherwise noted. Although $SiO_2$/Si substrates yielded flatter surfaces, they were not employed for optical measurements due to significant optical interference arising from multiple reflections[46]. Quartz substrates were cleaned using piranha solution, followed by UV-generated ozone treatment immediately prior to exfoliation. For growth, PTCDA powder (TCI, >98%) and the assembly templates were placed in the upstream and downstream zones of a three-zone tube furnace, respectively. After thoroughly purging the tube with Ar gas (99.999%) for 30 min at a flow rate of 500 mL/min, the flow rate was maintained at 100 mL/min throughout the growth process (20–40 min). The molecular source and template zones were maintained at 280–330 and 240–300 °C, respectively, with the source kept at a higher temperature. The thickness of PTCDA layers was determined with DR spectroscopy. The average thickness per growth batch was controlled by varying the growth time and temperature. Compared to the previous report using a single-zone furnace[8], the overall uniformity of PTCDA thickness was significantly improved, which was attributed to a reduced temperature gradient within the deposition area.

***AFM measurements***. The topographical details of the samples were characterized by an atomic force microscope (Park Systems Inc., XE-70). Height images were obtained in a non-contact mode using Si tips with a nominal radius of 8 nm (MicroMasch Inc., NSC-15).

***TEM measurements.*** To enhance stability under electron-beam irradiation, thin graphene layers were laminated onto the heterocrystals of PTCDA and 1L $MoS_2$[8]. Then, polydimethylsiloxane (PDMS) film (~0.6 mm thick), supported on a glass slide, was overlaid on the sandwich of graphene/PTCDA/$MoS_2$/quartz. While immersed in ethanol/DI water mixtures (ethanol 50%, DI water 50%) for 3 h, the stack was released from the quartz substrate and attached to the PDMS film. The PDMS-supported stack was dried under ambient conditions for 3–24 h. Subsequently, the stack was transferred onto holey SiN membrane TEM grids (Norcada Inc., hole diameter = 2 μm) under optical microscope observation using a conventional dry transfer method[47]. Selected area electron diffraction (SAED) measurements were obtained at 80 kV using a JEOL JEM-F200, with a typical selected probing diameter of 0.8 μm.

***PL spectroscopy.*** The micro-spectroscopy setup used in this study has been described elsewhere[42]. Briefly, excitation was provided by solid-state (514.3 nm; Cobolt, Fandango) and He-Ne (632.8 nm; Melles Griot) lasers, focused onto the samples to a spot size of ~1 μm using a microscope objective (40×, NA = 0.60). The emitted signals were collected using a spectrometer (Princeton Instruments, SP2300) equipped with a CCD camera (Princeton Instruments, PyLon). The overall spectral accuracy was better than 5 cm$^{-1}$. All measurements were performed under ambient conditions. The average power was maintained below 70 nW to minimize photoinduced degradation.

***PLE and DR spectroscopies***. The two measurements were performed with another micro-spectroscopy system[48], similar to the one described above, except for light sources. Each of the source beams was focused onto the sample, and signals collected in a backscattering geometry using a 40× objective lens (NA = 0.6) were then directed to a spectrometer equipped with a thermoelectrically cooled CCD detector (Andor Inc., DU971P). For PLE measurements, the excitation wavelength was tuned from 540 nm to 595 nm using an optical parametric oscillator (OPO; Coherent Inc., Chameleon Compact OPO-Vis) pumped by the Ti:sapphire laser (Coherent Inc., Chameleon Ultra II). The PL signals in the 640–730 nm range were integrated without polarization selection. The DR spectra were acquired using a collimated tungsten-halogen lamp as a broadband source in both unpolarized and polarization-resolved configurations, with the latter used to construct polarization-dependent intensity plots. The parallel-configuration DR ($DR^{\parallel}$) spectra were obtained by filtering the broadband source and reflected signals with linear polarizers. Samples were rotated using a rotational mount with a precision of 0.2°. The typical probe diameter was ~2 μm for DR measurements.

ASSOCIATED CONTENT

**Supporting Information**

Extraction of molecular absorption from DR of heterocrystals; DR spectra of pristine 1L $MoS_2$ ($1L_M$), annealed $1L_M$ and $1L_P/1L_M$ heterocrystal (HC); Polarization-dependent DR spectra of $2L_P/1L_M$; Orthogonality between the LDS and UDS components in 1–$5L_P/1L_M$; Spatial homogeneity of the optical response of $1L_P/1L_M$; Estimation of the excess electron density in $MoS_2$ of heterocrystals; Crystallographic parameters of $nL_P/1L_M$ heterocrystals; Number of samples examined for each optical measurement.

AUTHOR INFORMATION

**Corresponding Author**

*E-mail: sunryu@postech.ac.kr

NOTES

The authors declare no conflict of interest.

ACKNOWLEDGMENTS

This work was supported by the National Research Foundation of Korea (NRF-RS-2024-00336324, NRF-RS-2024-00411134, NRF-2021R1A6A1A10042944) and the Glocal University 30 project.

## Figure and Captions

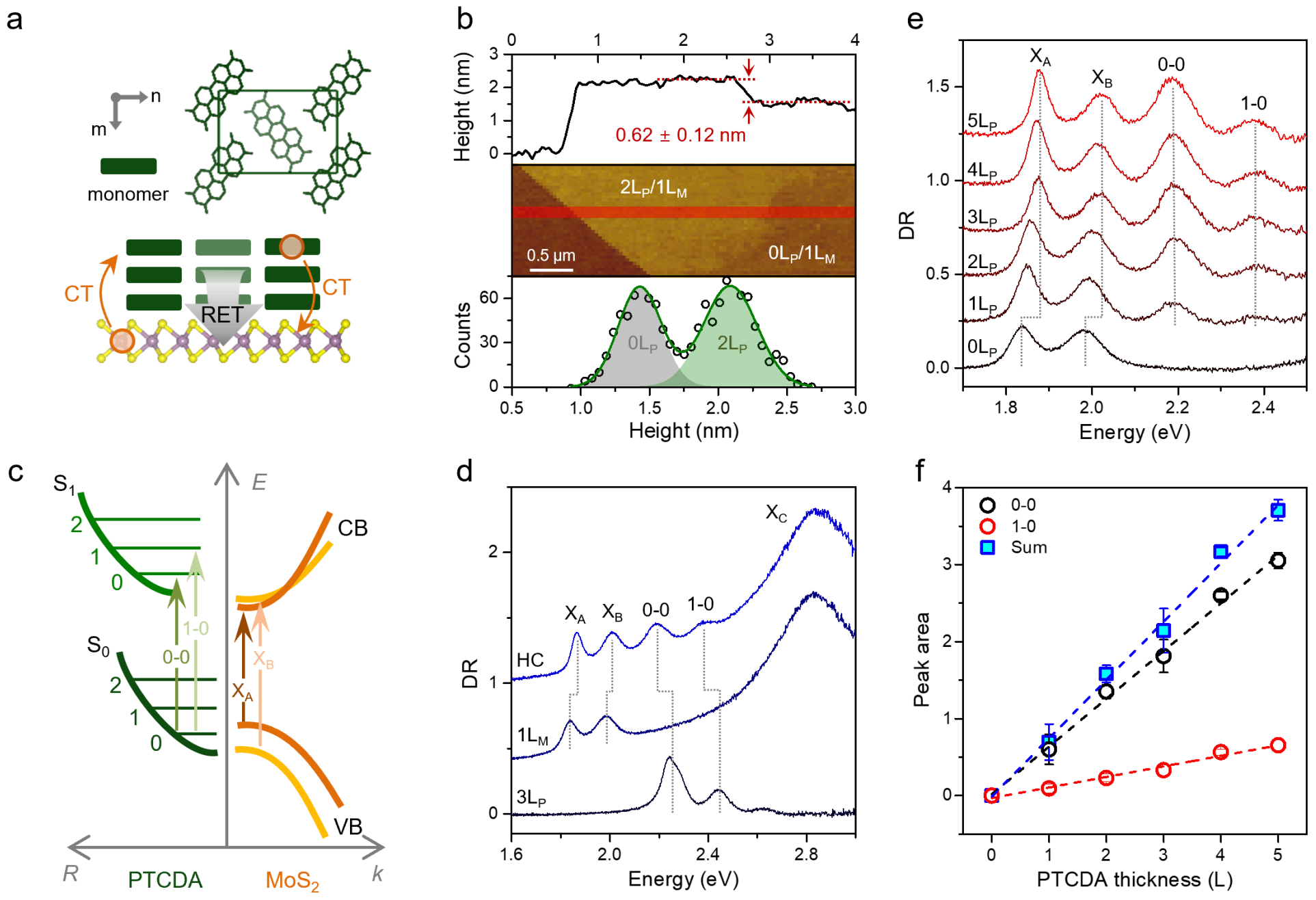

**Figure 1. Structural and optical characterization of thickness-controlled PTCDA/$MoS_2$ heterocrystals.** (a) Schematic diagram of the 2D molecular lattice (top) and side-view structure of the $3L_P/1L_M$ heterostructure (bottom), illustrating interfacial interactions via charge transfer (CT) and resonant energy transfer (RET). (b) AFM height image (middle) of $2L_P/1L_M$ with the height line profile (top) and height histogram (bottom). The profile was averaged over the red-shaded area in the image. (c) Schematic energy-level diagrams for PTCDA and monolayer $MoS_2$, qualitatively illustrating the molecular 0-0 and 1-0 vibronic transitions with the $X_A$ and $X_B$ excitons of $MoS_2$. (d) Unpolarized DR spectra of the $3L_P/1L_M$ heterocrystal (denoted as HC), monolayer $MoS_2$ ($1L_M$) and trilayer PTCDA on hBN ($3L_P$). (e) Baseline-subtracted unpolarized DR spectra of 0–$5L_P/1L_M$. The DR spectra of HCs were fitted using 5 Gaussian functions and a quadratic baseline (see Fig. S1 for details). (f) Peak areas of 0-0 and 1-0 molecular absorption and their sum as a function of PTCDA thickness.

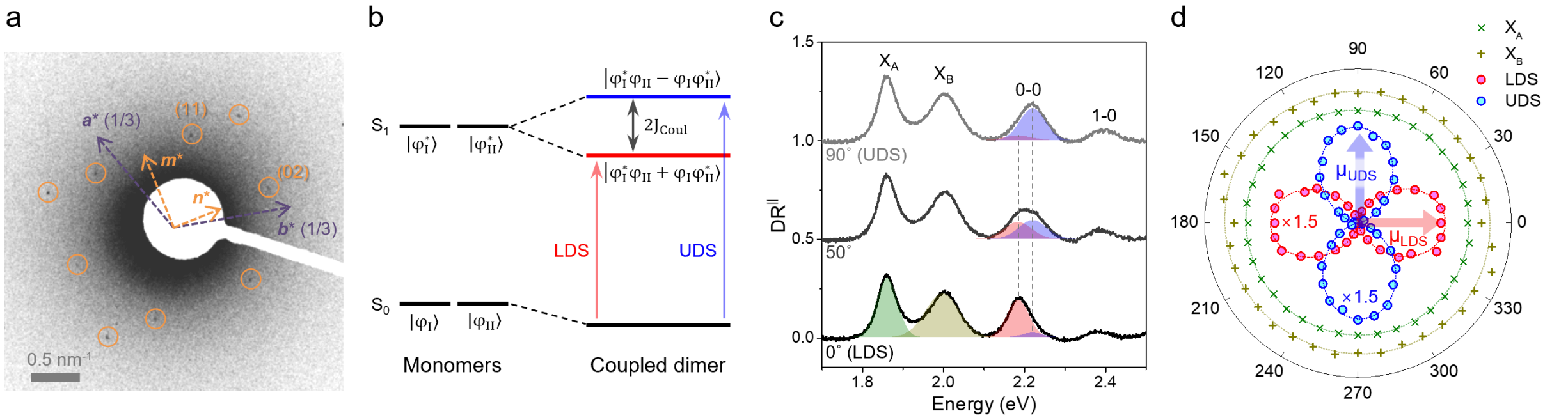

**Figure 2. Epitaxial registry and optical anisotropy of PTCDA/MoS₂ heterocrystals**. (a) Representative SAED pattern of graphene-encapsulated $2L_P/1L_M$, showing the reciprocal primitive vectors of PTCDA (***m****** and ***n****) and $MoS_2$ (***a**** and ***b****). The latter are reduced in size threefold. See Table S1 for quantitative analysis and statistics. (b) Exciton splitting in a coupled molecular dimer, illustrating the formation of the lower and upper Davydov states (LDS and UDS). $|\varphi_j>$ and $|\varphi_j^*>$ represent the ground and excited states of the j-th molecule, respectively. (c) Polarization-resolved parallel-configuration differential reflectance ($DR^{\parallel}$) spectra of $2L_P/1L_M$ at representative polarization angles shown with baseline subtracted. The full-set spectra at all angles (Fig. S3a) were globally fitted using five Gaussian functions (colored shades) to represent the $X_A$, $X_B$, LDS, UDS and 1-0 transitions. (d) Polar graph of $DR^{\parallel}$ areas for the LDS, UDS, $X_A$ and $X_B$ transitions, showing the orthogonality between the two Davydov states represented by the transition dipole moments ($\mu_{LDS}$ and $\mu_{UDS}$).

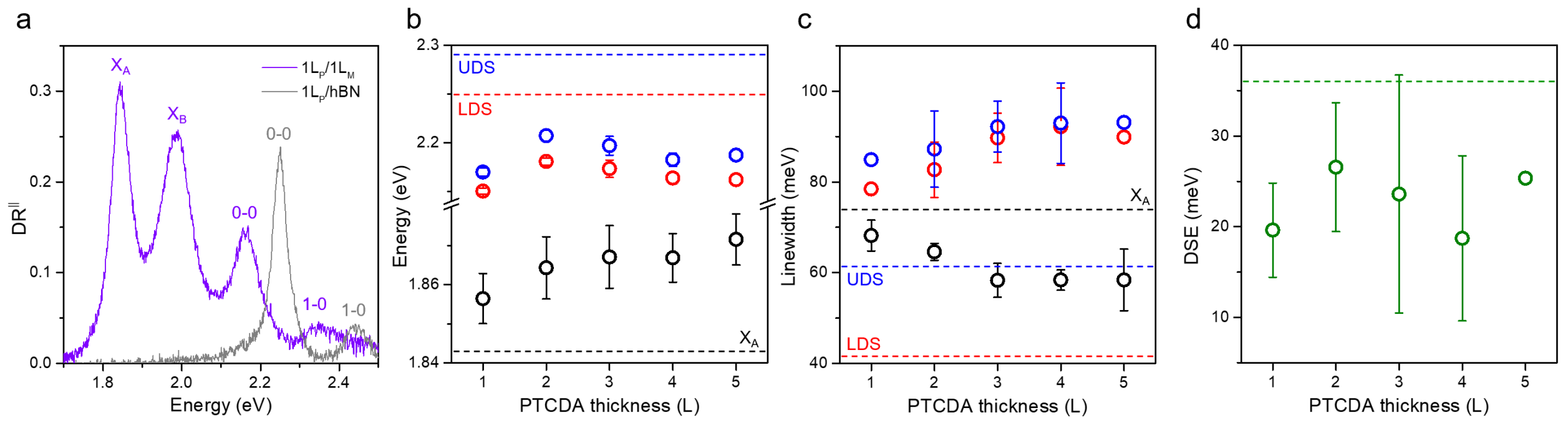

**Figure 3. Static interlayer coupling in PTCDA/$MoS_2$ heterocrystals.** (a) $DR^{\parallel}$ spectra of $1L_P/1L_M$ and the reference $1L_P$/hBN. (b) Thickness dependence of the LDS, UDS and $X_A$ energies. (c) Thickness dependence of the spectral linewidth (FWHM) of the LDS, UDS and $X_A$ resonances. (d) Davydov splitting energy (DSE) as a function of PTCDA thickness. The horizontal dashed lines in (b–d) indicate the corresponding reference energies. Error bars represent the standard deviation across multiple heterocrystals when applicable.

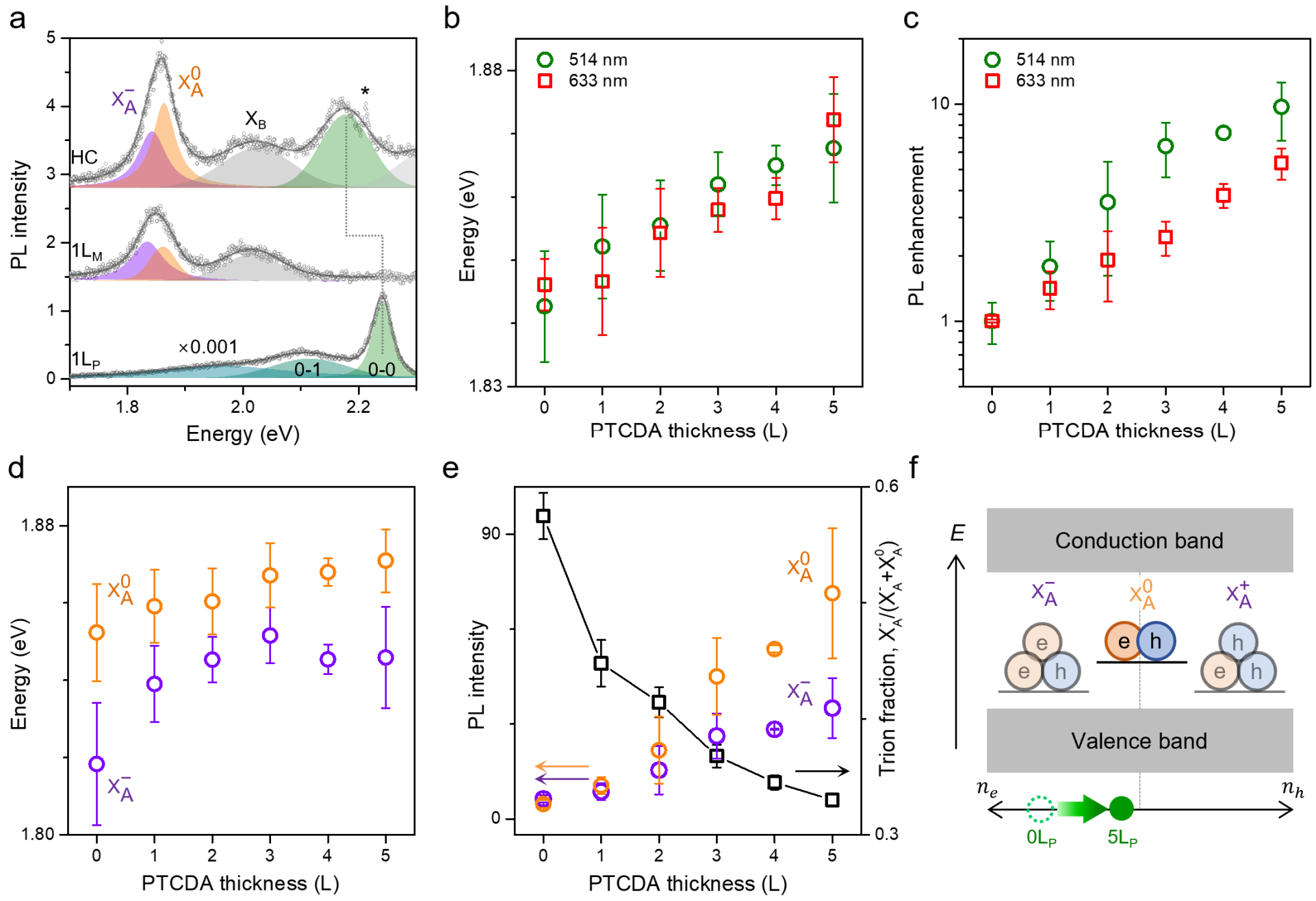

**Figure 4. Photoluminescence modulation and ground-state charge-transfer doping.** (a) Steady-state PL spectra of $1L_P/1L_M$ with bare $1L_M$ (annealed) and $1L_P$/hBN reference samples. $X_A$ is decomposed into two Voigt functions, representing the neutral exciton ($X_A^0$) and negative trion ($X_A^-$). PTCDA's Raman peak is marked with an asterisk. (b) Thickness dependence of the $X_A$ energy under above-gap (514 nm) and sub-gap (633 nm) excitations. (c) Thickness dependence of the PL enhancement of $X_A$ for the two excitation wavelengths. (d) Energies of the $X_A^0$ and $X_A^-$ components extracted from spectral fitting. (e) Thickness dependence of the PL intensities, $X_A^0$ and $X_A^-$, and the trion fraction in intensity, $X_A^-/(X_A^- + X_A^0)$. (f) Schematic illustration of ground-state charge-transfer doping, showing electron transfer from $MoS_2$ and the resulting conversion of $X_A^-$ to $X_A^0$. Error bars represent the standard deviation across multiple heterocrystals when applicable.

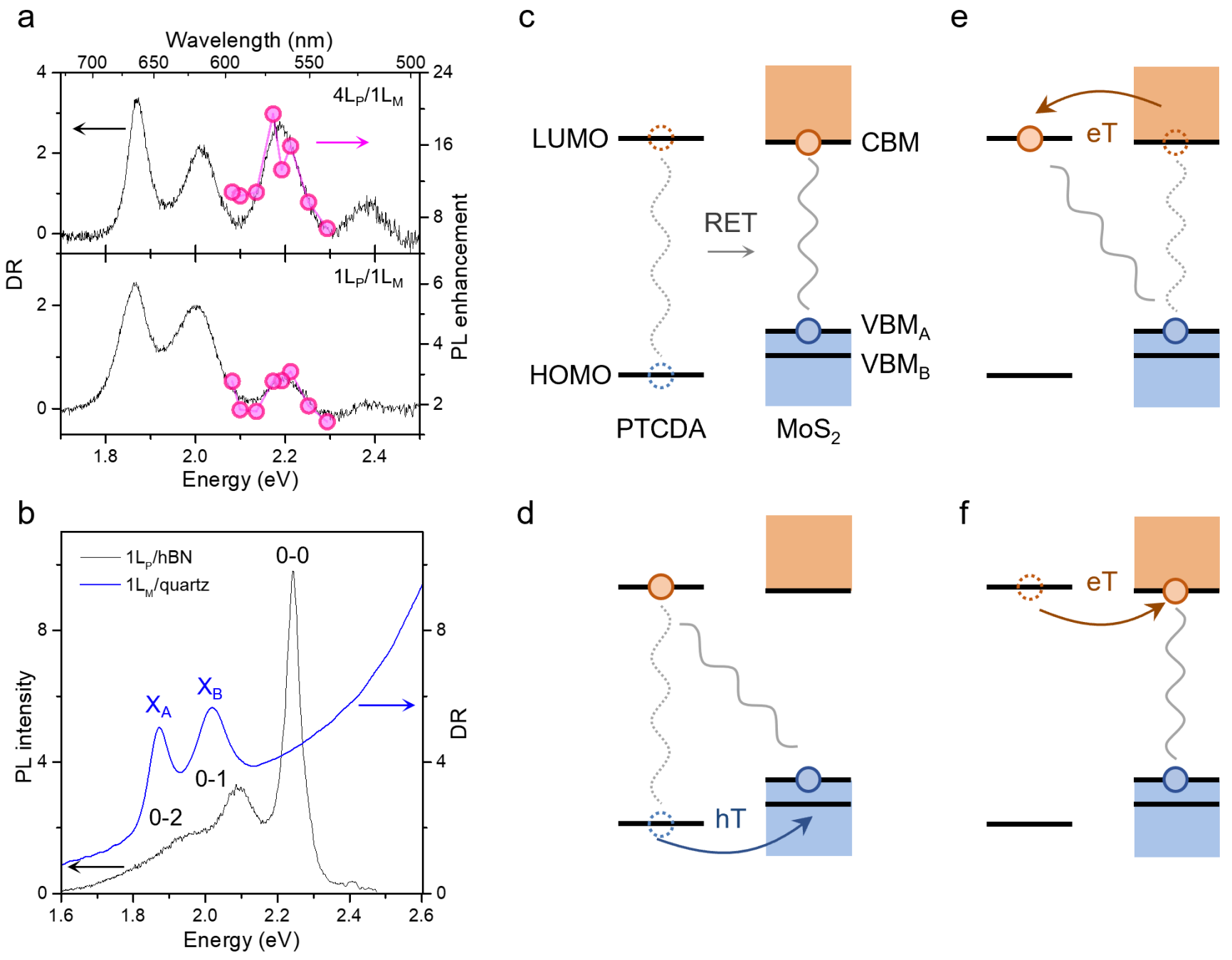

**Figure 5. Dynamic interlayer coupling and exciton transfer pathways.** (a) PLE spectra of $1L_P/1L_M$ and $4L_P/1L_M$, plotted with the corresponding DR spectra. (b) Spectral overlap between PTCDA's PL ($1L_P$/hBN) and $MoS_2$'s absorption represented by DR ($1L_M$/quartz). (c–d) Schematic energy-level diagrams qualitatively illustrating above-gap excitation: RET from photoexcited PTCDA to $MoS_2$ (c) and hole transfer (hT) from PTCDA to $MoS_2$ (d). (e–f) Schematic energy-level diagrams illustrating sub-gap excitation: forward electron transfer (eT) from photoexcited $MoS_2$ to PTCDA (e), followed by possible back electron transfer to $MoS_2$ (f). The processes in (e–f) may occur for above-gap excitation.

**TOC Figure**

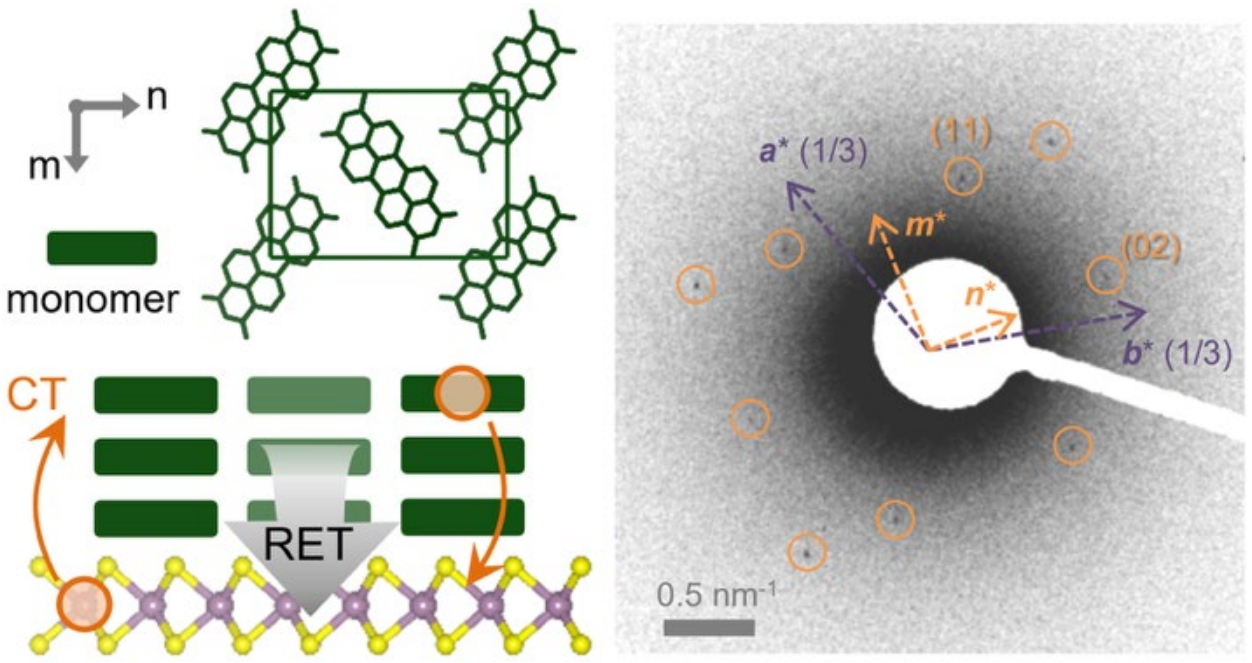

Supplementary Information for

# Dual-Mode Exciton Coupling in Epitaxially Registered Organic-Inorganic 2D Heterocrystals

Eunbeen Jeon, Kihyun Lee, Jieun Yeon, Juseung Oh, Kenji Watanabe, Takashi Taniguchi, Kwanpyo Kim and Sunmin Ryu*

*E-mail: sunryu@postech.ac.kr

**Contents**

## Supplementary Figures

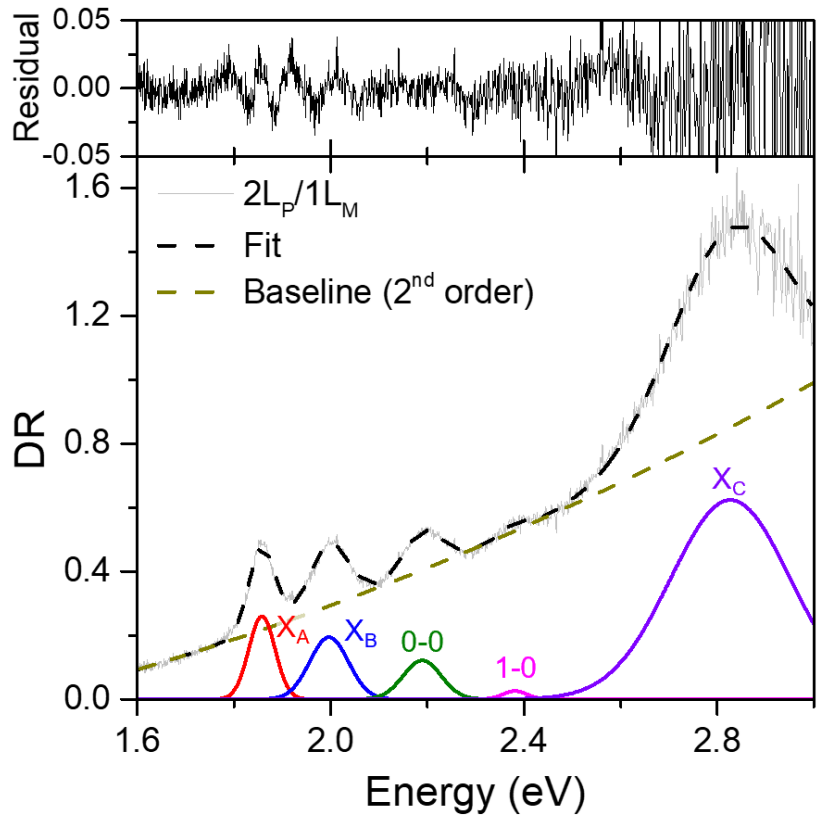

**Figure S1. Extraction of molecular absorption from DR of heterocrystals.** Raw DR spectra were fitted with multiple Gaussian and n$^{th}$-order polynomial (n = 1–3, dashed line) functions. The overall goodness of fit can be judged from the small residual, which was not affected meaningfully by the choice of n. Whereas the peak areas of 0-0 and $X_A$ also remained within 8% for n = 1–3, a quadratic baseline function, also used for $WS_2$[1], was adopted due to slightly better fitting reliability.

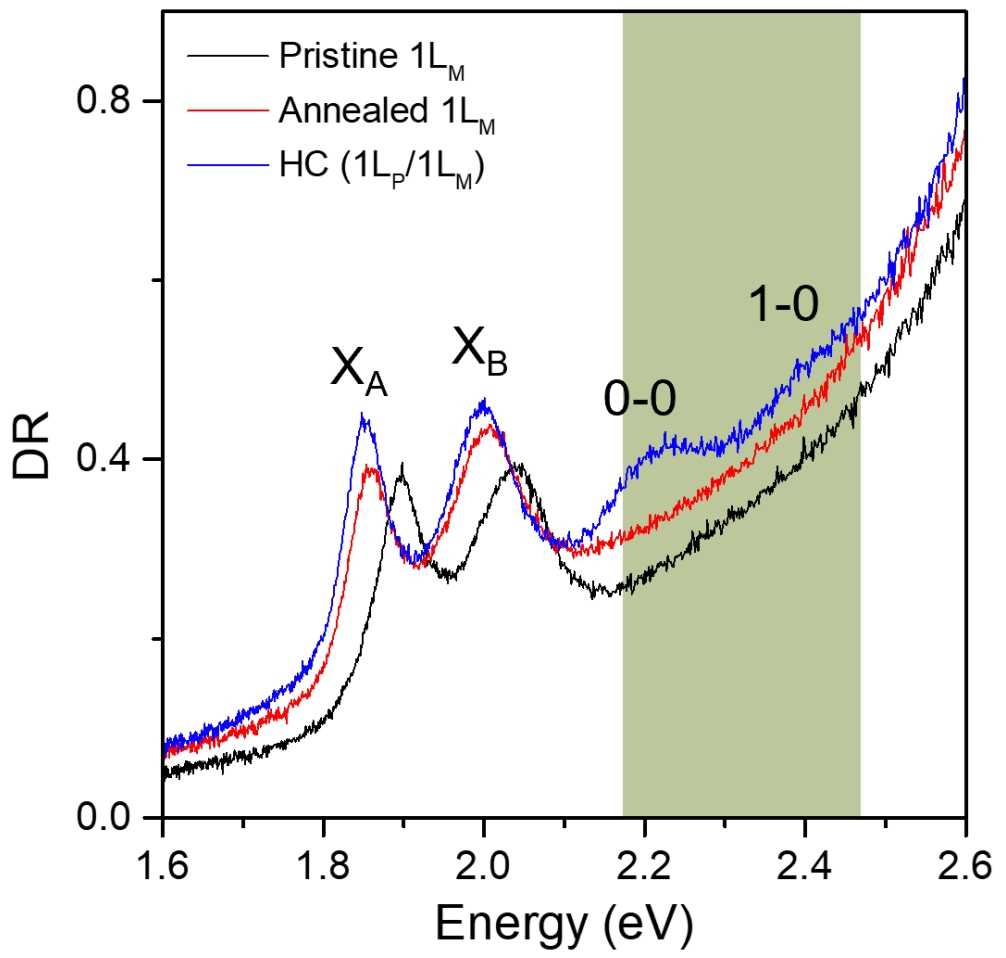

**Figure S2. DR spectra of pristine 1L $MoS_2$ (1L$_M$), annealed 1L$_M$ and 1L$_P$/1L$_M$ heterocrystal (HC).** Upon annealing, the DR signals of $MoS_2$ change significantly, particularly over the energy range of the molecular 0-0 and 1-0 absorption (denoted by the shaded region). As a result, the conventional method of determining the thickness of PTCDA layers from optical contrast[2] is unreliable.

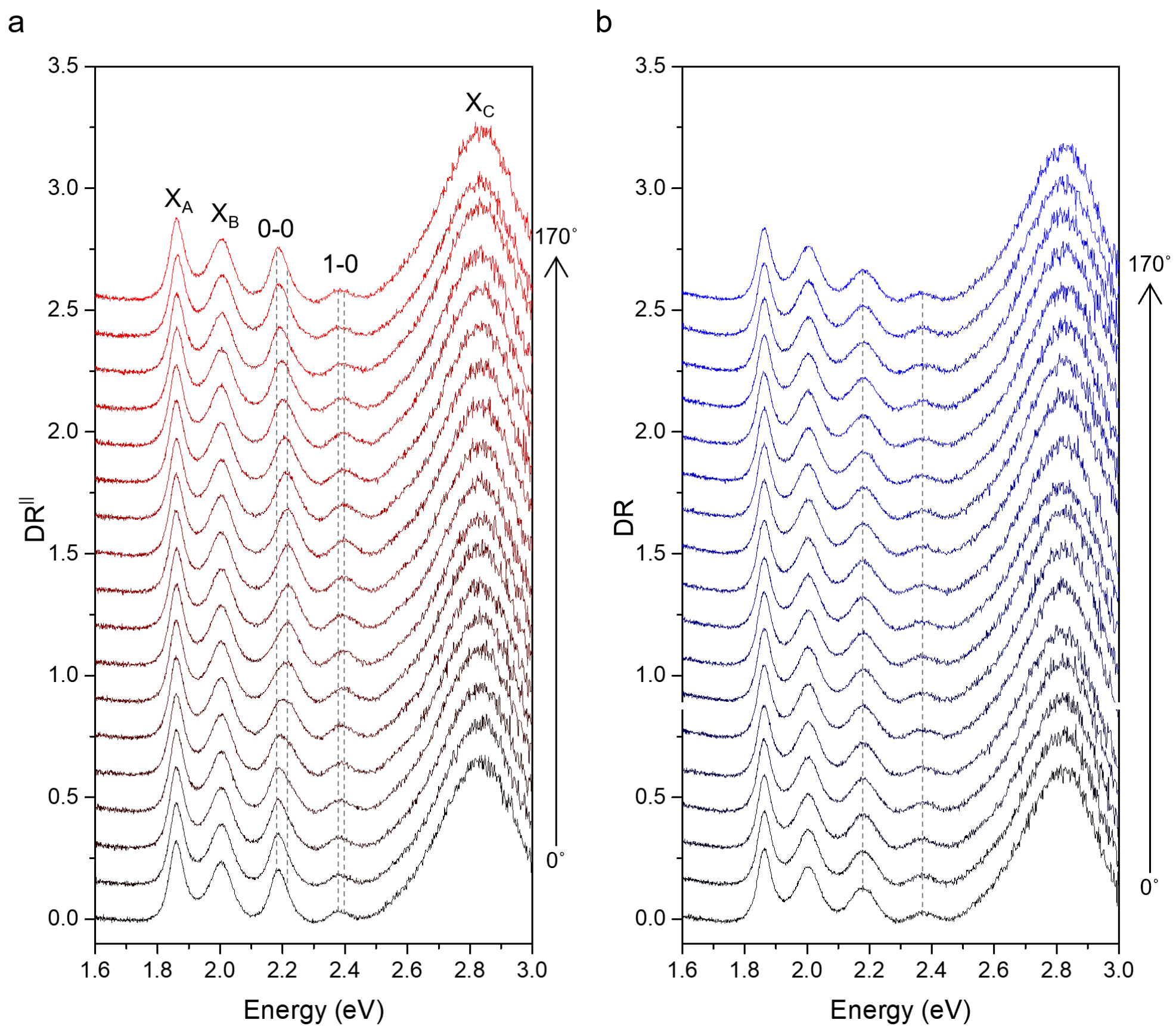

**Figure S3. Polarization-dependent DR spectra of 2Lp/1Lm.** (a–b) Polarization-resolved parallel-configuration DR (DR$^{\parallel}$) spectra (a) and unpolarized DR spectra (b), acquired at 10° intervals from 0° to 170°. Vertical dashed lines indicate the LDS and UDS components of the 0-0 and 1-0 transitions in (a) and the unresolved 0-0 and 1-0 transitions in (b).

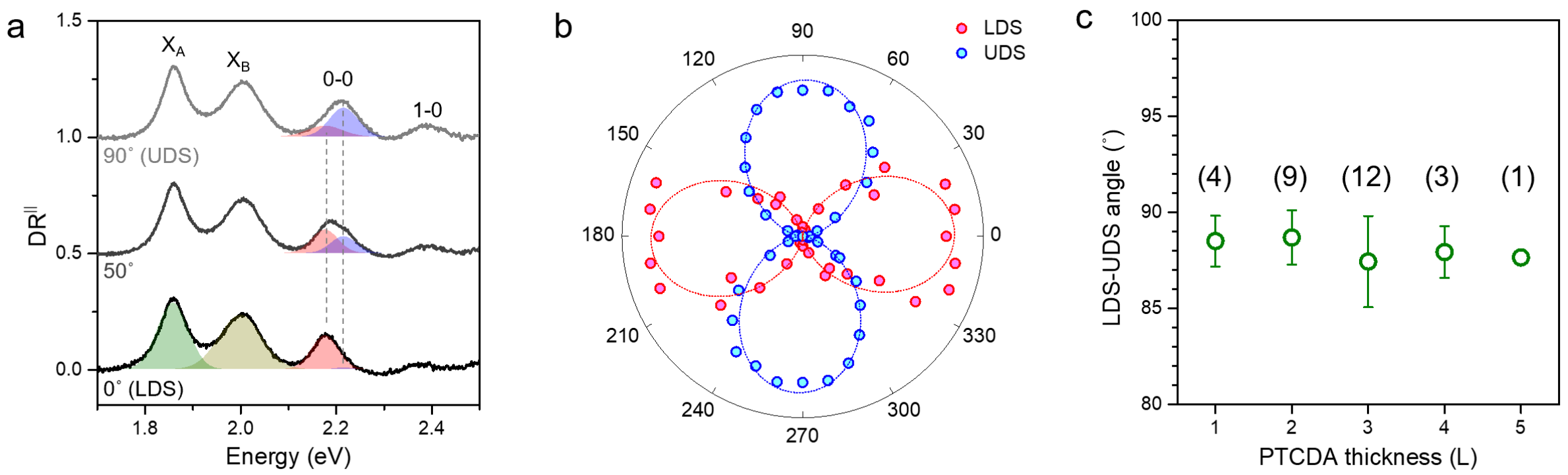

**Figure S4. Orthogonality between the LDS and UDS components in 1–5L$_P$/1L$_M$.** (a) Polarization-resolved parallel-configuration differential reflectance (DR$^{\parallel}$) spectra of 1L$_P$/1L$_M$ at representative polarization angles shown with baseline subtracted. The spectra were globally fitted using five Gaussian functions (colored shades) to represent the X$_A$, X$_B$, LDS, UDS and 1-0 transitions. (b) Polar graph of DR$^{\parallel}$ areas of the LDS and UDS components for 1L$_P$/1L$_M$, as in Fig. 2d for 2L$_P$/1L$_M$. (c) Angles between the LDS and UDS dipoles, defined between 0 and 90°, as a function of PTCDA thickness. The numbers in parentheses indicate the number of samples measured at each thickness. The mean angle across all 29 samples is 88.0 ± 1.9°.

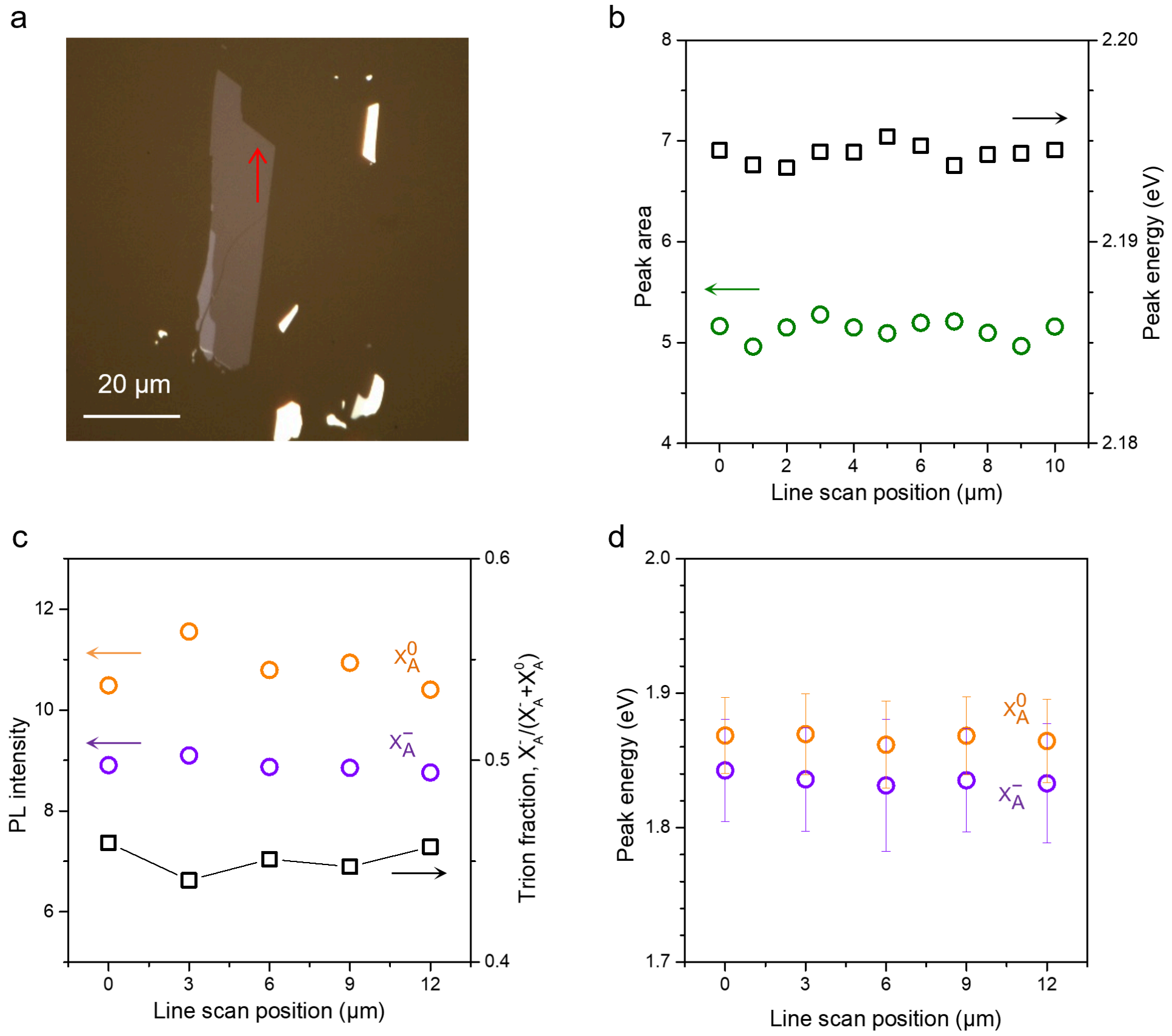

**Figure S5. Spatial homogeneity of the optical response of 1LP/1LM.** (a) Optical micrograph showing the line-scan direction (red arrow). (b) PTCDA 0-0 peak area and peak energy obtained from DR spectra along the line scan, with average values of 5.1 ± 0.1 (RSD = 2%) and 2.194 ± 0.001 eV (RSD = 0.05%), respectively. (c) $MoS_2$ neutral-exciton ($X_A^0$) and trion ($X_A^-$) PL intensities obtained from Voigt fitting, together with the trion fraction. The average trion fraction is 0.451 ± 0.008 (RSD = 2%). (d) Peak energies of $X_A^0$ and $X_A^-$ along the line scan. The vertical bars represent the corresponding fitted linewidths. The average peak energies are 1.866 ± 0.003 eV (RSD = 0.2%) and 1.836 ± 0.004 eV (RSD = 0.2%) for $X_A^0$ and $X_A^-$, respectively.

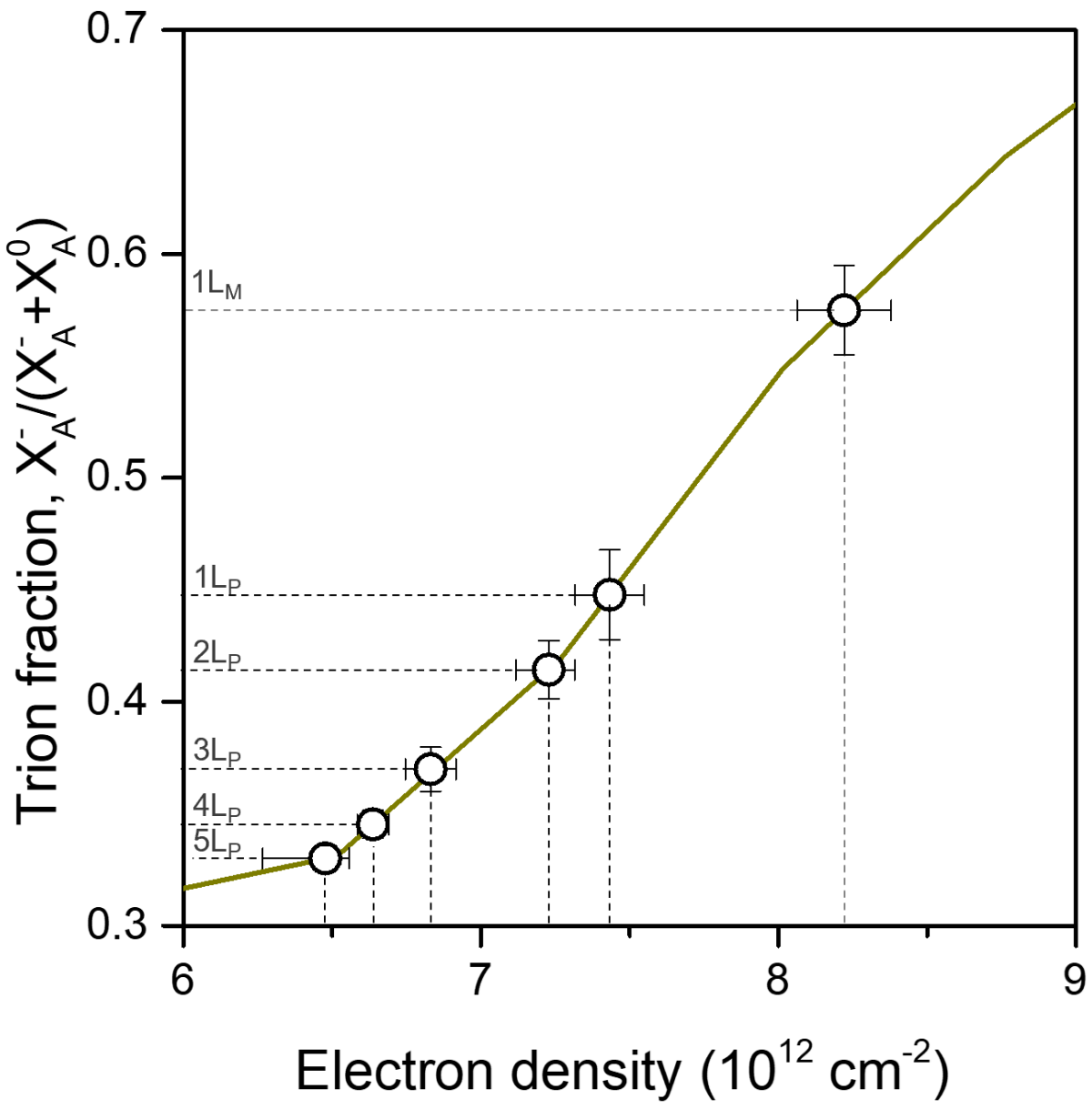

**Figure S6. Estimation of the excess electron density in $MoS_2$ of heterocrystals.** From PL measurements[3] of electrically doped 1L $MoS_2$, the trion fraction in the PL of the A exciton, $X_A^-/(X_A^- + X_A^0)$, is plotted as a function of the effective electron density in 1L $MoS_2$ (solid line). The conversion of trion fraction to electron density assumes that the relationship established for electrically gated monolayer $MoS_2$ remains approximately applicable to the present PTCDA/$MoS_2$ heterocrystal systems. Thus, the resulting electron densities should be regarded as approximate estimates. The open circles denote the average trion fractions measured for $nL_P/1L_M$ (Fig. 4e) and their corresponding electron densities. Vertical error bars represent the standard deviation of the trion fraction measured across independent heterocrystals. Horizontal error bars show the corresponding electron-density ranges obtained by converting the mean trion fraction ± one standard deviation through the calibration curve. Their asymmetry results from the trion fraction's nonlinear dependence on electron density.

## Supplementary Table

**Table S1. Crystallographic parameters of $nL_P/1L_M$ heterocrystals.** θ denotes the angle between two vectors. The crystallographic parameters were obtained from multiple spots in two prepared $2L_P$ samples and one spot in one $4L_P$ sample. Values in parentheses and square brackets indicate one standard deviation in the last digits and the percentage deviation from bulk PTCDA[4], respectively.

| Samples | \|***m***\| (nm) | \|***n***\| (nm) | \|***n***\|/\|***m***\| | θ(***m***-***n***) (˚) | θ(***b***-**n****) (˚) | Note |
|---|---|---|---|---|---|---|
| $2L_P/1L_M$ | 1.276(22) | 1.971(57) | 1.544(22) | 89.3(4) | 12.7(4) | |
| | [+6.7%] | [-1.0%] | [-7.3%] | [-0.8%] | | |
| $4L_P/1L_M$ | 1.224 | 1.992 | 1.628 | 91.7 | 12.7 | |
| | [+2.3%] | [+0.1%] | [-2.2%] | [+1.9%] | | |
| $1L_P$/hBN | 1.246 | 1.991 | 1.598 | 89.5 | - | Ref. 2 |
| | [+4.2%] | [0%] | [-4.0%] | [-0.6%] | | |
| $3L_P$/hBN | 1.186 | 1.981 | 1.670 | 89.4 | | |
| | [-0.8%] | [-0.5%] | [+0.3%] | [-0.7%] | | |
| Bulk crystals | 1.196 | 1.991 | 1.665 | 90 | - | Ref. 4 |

**Table S2. Number of samples examined for each optical measurement.**

| Measurement | PTCDA thickness | | | | | |
|---|---|---|---|---|---|---|
| | 0L (bare $MoS_2$) | 1L | 2L | 3L | 4L | 5L |
| DR | 1 | 10 | 10 | 7 | 5 | 8 |
| DR$^{\parallel}$ | 1 | 4 | 9 | 12 | 3 | 1 |
| PL (514 nm) | 4 | 5 | 3 | 7 | 2 | 3 |
| PL (633 nm) | 4 | 4 | 3 | 7 | 3 | 3 |
| PLE | 1 | 1 | - | - | 1 | - |

**Supplementary Note**

**Minimal kinetic framework for interfacial excitation pathways**

To organize the interfacial excitation pathways, we consider a minimal population model containing the photoexcited PTCDA population ($P^*$), the emissive $MoS_2$ exciton population ($X$) and a photoinduced interfacial charge-transfer population ($C$). Interfacial charge- and energy-transfer pathways of this type have been reported for PTCDA/$MoS_2$ and related organic/TMD heterostructures[5, 6]. The ground-state charge-transfer doping discussed in the main text is distinct from these excited-state pathways: It changes the equilibrium carrier density of MoS2 and thereby its exciton-trion population balance and effective radiative response, without requiring photoexcitation of PTCDA.

$$\frac{dP^*}{dt} = G_P - (k_{P,r} + k_{P,nr} + k_{RET} + k_{CT})P^*$$

$$\frac{dX}{dt} = G_M + \eta_{RET} k_{RET} P^* - \left(k_{X,r} + k_{X,nr}\right)X$$

$$\frac{dC}{dt} = k_{CT} P^* - k_C C$$

, where $G_P$ and $G_M$ denote the generation rates of photoexcited PTCDA and $MoS_2$ excitons, respectively. $k_{P,r}$ and $k_{P,nr}$ are the intrinsic radiative and nonradiative decay rates of photoexcited PTCDA, while $k_{X,r}$ and $k_{X,nr}$ are the corresponding rates for the emissive $MoS_2$ exciton population. $k_{RET}$ and $k_{CT}$ denote resonance energy transfer and photoinduced interfacial charge transfer from excited PTCDA, respectively. $\eta_{RET}$ represents the fraction of RET events that populate the emissive $MoS_2$ exciton manifold, whereas $k_C$ denotes the effective decay rate of the interfacial charge-transfer population, encompassing all subsequent pathways that depopulate $C$.

Under steady-state excitation,

$$P^* = \frac{G_P}{k_{P,r} + k_{P,nr} + k_{RET} + k_{CT}},$$

$$X = \frac{G_M + \eta_{RET} k_{RET} P^*}{k_{X,r} + k_{X,nr}},$$

and

$$C = \frac{k_{CT} P^*}{k_C}.$$

The corresponding molecular ($I_P$) and $MoS_2$ ($I_M$) PL intensities can be written as

$$I_P \propto \frac{k_{P,r} G_P}{k_{P,r} + k_{P,nr} + k_{RET} + k_{CT}},$$

$$I_M \propto \frac{k_{X,r}}{k_{X,r} + k_{X,nr}} \left[ G_M + \eta_{RET} k_{RET} \frac{G_P}{k_{P,r} + k_{P,nr} + k_{RET} + k_{CT}} \right].$$

The steady-state CT population, $C = \frac{k_{CT} P^*}{k_C}$, describes the population generated through photoinduced charge transfer and subsequently depleted through effective CT-state decay. Because $C$ is treated here as a downstream population without feedback into $P^*$ or $X$, $k_C$ does not enter the expressions for the molecular or $MoS_2$ PL intensities.

This framework also clarifies the distinction between the two experimental excitation conditions. Under 633 nm excitation, direct PTCDA excitation is negligible ($G_P \approx 0$), so the molecular-excitation-dependent terms involving $P^*$ vanish. PTCDA-dependent changes in the $MoS_2$ PL can nevertheless remain through the ground-state charge-transfer doping, which modifies the exciton-trion population balance and effective radiative and nonradiative decay rates of $MoS_2$.

Under 514 nm excitation, $G_P \neq 0$, and photoexcited PTCDA introduces additional interfacial pathways. In particular, the RET contribution to the $MoS_2$ exciton population scales with

$$\eta_{RET} \frac{k_{RET}}{k_{P,r} + k_{P,nr} + k_{RET} + k_{CT}}.$$

Thus, RET competes directly with photoinduced charge transfer and the intrinsic radiative and nonradiative decay channels of PTCDA for the same molecular excited-state population. The strong quenching of PTCDA PL indicates that such interfacial pathways substantially increase the decay of $P^*$.

The present steady-state observables, however, do not independently determine $k_{RET}$, $k_{CT}$, $k_C$ or the competing nonradiative rates. RET and photoinduced charge transfer both depopulate $P^*$, so the present steady-state PL and PLE measurements do not determine $k_{RET}$ and $k_{CT}$ independently. Separating these competing pathways quantitatively would require time-resolved measurements.

## Supplementary References